\documentstyle[11pt,newpasp,twoside,psfig]{article}

\begin{document}

\title{Diffraction Limited Imaging of High Redshift Galaxies with Adaptive Optics}
\author{R. I. Davies, M. Lehnert, A. J. Baker, and S. Rabien}
\affil{Max-Planck-Institut f\"ur extraterrestrische Physik, Postfach 1312, 85741 Garching, Germany}

\begin{abstract}
The major cornerstone of future ground-based astronomy is imaging
and spectroscopy at the diffraction limit using adaptive optics.
To exploit the potential of current AO systems, we have begun a survey 
around bright stars to study intermediate redshift galaxies at high 
resolution. 
Using ALFA to reach the diffraction limit of the 3.5-m telescope
at Calar Alto allows us to study the structure of distant galaxies in the 
near-infrared at scales of 100--150\,pc for $z$=0.05 and at scales
1.0--1.5\,kpc at $z$=1.
In this contribution we present the initial results of this project,
which hint at the exciting prospects possible with the resolution and
sensitivity available using an AO camera on the 8-m class VLT.
\end{abstract}

\smallskip

Resolution and sensitivity are crucial for studies at high redshift of
galaxy dynamics and demographics, quasar hosts, etc.
The immense potential of modern telescopes for such work is largely
wasted unless adaptive optics (AO) can be used to correct for the seeing
induced by atmospheric turbulence.
Since it is likely to be several years before laser guide stars begin
to be used as a standard tool, we must observe in the vicinity
of natural guide stars.
But all current deep surveys deliberately avoid stars of any sort,
leaving the AO without a reference.
Instead we can invert the problem and look for target objects near the
best reference stars, a technique that we and Larkin \& Glassman (1999)
have shown to be successful.

\smallskip

For AO to make an impact on astrophysics, a Strehl ratio of 20\% can
be considered a realistic minimum correction for the K-band.
But, because any target object is likely to be
20--30$^{\prime\prime}$ from a reference star, to achieve this we
require 40--50\% Strehl on the star (Le~Louarn et al. 1999).
The star must therefore be bright, $V<12$ (depending on the wavefront
sensor).
To reduce scattered light in the infrared bands we can choose to search near
only early type stars -- the scattered K-band light from a 12\,mag A-star
is nearly the same as that from a randomly chosen 14\,mag star because
most of these are G-type or later.
Lastly, selecting stars away from the Galactic plane minimises both
the number of background stars and the extinction.

\smallskip

We have made a 1\,hr exposure in the K$^\prime$-band at the 3.5-m telescope
at Calar Alto (Spain), with the 80$^{\prime\prime}$ field centered on
the $V=9.08$ A0 star SAO\,81538.
The MPIA/MPE adaptive optics system ALFA (Kasper et al. 2000) provided
a Strehl ratio of 32\% in seeing of 0.87$^{\prime\prime}$.
The high Strehl makes the data particularly sensitive to point sources
-- i.e., stars, or galaxies with bright compact nuclei such as AGN or
starbursts.
On the other hand, the small pixel scale means that faint extended
objects might be missed.
The caveat is that the population of objects seen in such an image is
almost certainly different from, and a subset of, those generally
investigated in seeing-limited data.

\smallskip

\begin{figure}
\centerline{\psfig{file=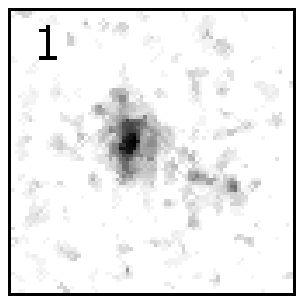,width=3cm}\hspace{2mm}\psfig{file=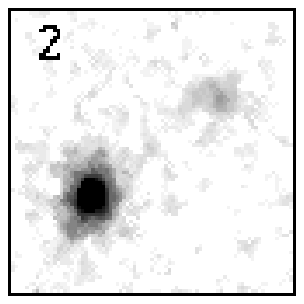,width=3cm}\hspace{2mm}\psfig{file=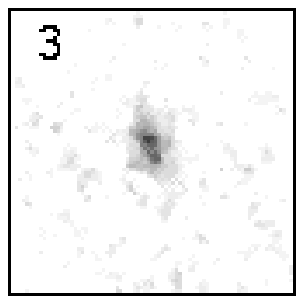,width=3cm}\hspace{2mm}\psfig{file=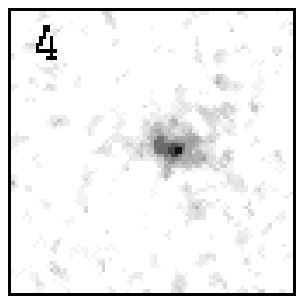,width=3cm}}
\vspace{2mm}
\centerline{\psfig{file=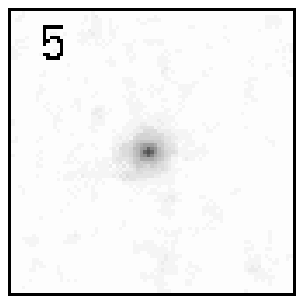,width=3cm}\hspace{2mm}\psfig{file=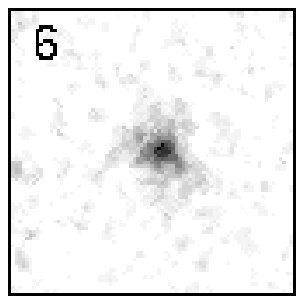,width=3cm}\hspace{2mm}\psfig{file=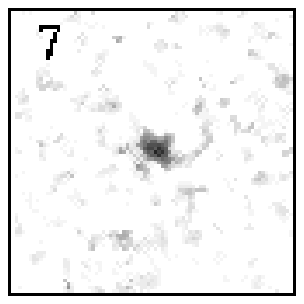,width=3cm}\hspace{2mm}\psfig{file=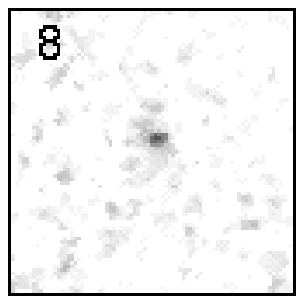,width=3cm}}
\caption{6.5$^{\prime\prime}$$\times$6.5$^{\prime\prime}$ cutouts of
objects within 30$^{\prime\prime}$ of the guide star,
smoothed with a median filter of radius 0.16$^{\prime\prime}$.
Upper row: interacting and multiple-nuclei galaxies.
Lower row: galaxies with single nuclei.}
\label{cutouts}
\end{figure}

\begin{figure}[h]
\begin{minipage}[b]{6.8cm}
The cutouts in Fig.~\ref{cutouts} show that we have detected at least
8 galaxies, 4 of which are interacting or have multiple nuclei.
All the objects in the figure have magnitudes K$\sim$19--20, for
which we estimate a mean redshift of 0.7 (Cowie et al. 1996).
The structure of the 4 faintest objects (not
shown), which have K=20.5--20.7, cannot be determined from these data
alone.

\smallskip
The graph in Fig.~\ref{profiles} shows the radial
profiles for the on-axis reference star and 3 objects similar
distances from it:
another star which defines the PSF, and 2 single nuclei galaxies
(objects~5 \&~6 above).
The core of the galaxy which is object~6 is resolved even though it
has an intrinsic {\sc FWHM}$<$0.2$^{\prime\prime}$.
\end{minipage}
\begin{minipage}[b]{6.8cm}
\psfig{file=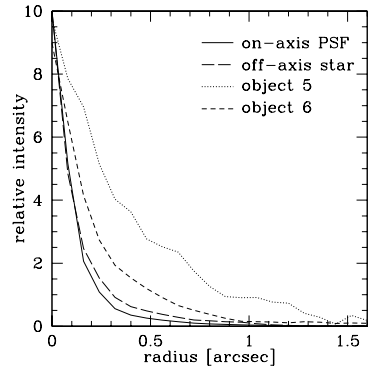,height=5cm}
\caption{Azimuthally averaged profiles for the on-axis reference star
and 3 objects all similar distances from it.}
\label{profiles}
\end{minipage}
\end{figure}

\end{document}